\documentstyle[art12]{article}
\def\al{\mbox{$\alpha_S$}}
\def\nn{\nonumber\\}

\begin{document}
\begin{titlepage}
\begin{flushright}
FERMILAB-Pub-95/127-T\\
DTP/95/52\\
\end{flushright}
\vspace{1cm}
\begin{center}
{\Large\bf The Determination of $\al$ at Hadron Colliders}\\
\vspace{1cm}

{\large
W.~T.~Giele}\\
\vspace{0.4cm}
{\it
Fermi National Accelerator Laboratory, P.~O.~Box 500,\\
Batavia, IL 60510, U.S.A.} \\
\vspace{0.4cm}
{\large
E.~W.~N.~ Glover}\\
\vspace{0.4cm}
{\it
Physics Department, University of Durham,\\ Durham DH1~3LE, England}
\\
\vspace{0.4cm}
and \\
\vspace{0.4cm}
{\large
J.~ Yu}\\
\vspace{0.4cm}
{\it
Department of Physics and Astronomy, University of Rochester\\

Rochester, NY 14627, U.S.A.} \\
\vspace{0.4cm}
{\large \today}
\vspace{0.4cm}
\end{center}

\begin{abstract}
Hadron colliders offer a unique opportunity

to test perturbative QCD because, rather than

producing events at a specific beam energy, the dynamics

of the hard scattering is probed simultaneously at a wide

range of momentum transfers.

This makes the determination of $\al$

and the parton density functions (PDF) at
hadron colliders particularly interesting.

In this paper we restrict ourselves
to extracting $\al$ for a given PDF at a scale

which is directly related

to the transverse energy produced in the collision.
As an example, we focus on

the single jet inclusive transverse energy

distribution and use the published '88-'89 CDF data
with an integrated luminosity of 4.2 pb$^{-1}$.
The evolution of the coupling constant over a wide
range of scales (from 30~GeV to 500~GeV) is clearly shown and

is in agreement with the QCD expectation.
The data to be obtained in the current
Tevatron run (expected to be well in excess 100 pb$^{-1}$  for both
the CDF

and D\O\ experiments) will significantly decrease the experimental
errors.
\end{abstract}

\end{titlepage}

\section{Introduction}

Hadronic collisions at the  Fermilab TEVATRON

offer excellent opportunities to study QCD over a

broad range of momentum transfers ranging from a few GeV in the
transverse

momentum distribution of the $Z$ boson
up to almost half of the beam energy in the single jet

inclusive transverse energy distribution.
While the experiments at LEP and HERA have

set well defined goals for QCD studies,

hadron colliders tend to be thought of as
discovery machines probing the high energy frontier.
For example, at Fermilab, the major effort has been concentrated
on the study of the Top Quark and $W$-mass measurements.

In this paper, we try to redress this imbalance and outline a
possible goal for QCD studies at hadron colliders.

Achieving this goal will give both a rigorous

test of QCD and a reduction of the experimental systematic errors
in the other studies at Fermilab.

One possible goal of QCD studies at the Main Injector \cite{maininj}
is
to use the QCD data set to
determine the input parameters of the theory,

in other words,  $\al$ and the parton density functions (PDF's),

{\em without} input
from other experiments.\footnote{Note that the range of $x$ and $Q^2$

probed in hadron-hadron collisions is rather different from that
probed at HERA.}
This should also allow the determination of the gauge
symmetry responsible for the strong interactions thereby
extending similar measurements at LEP \cite{LEPgauge}.

In order to achieve this goal,
we need to make several intermediate steps to identify problems in
both

experiments and theory. The run 1A and 1B data
can be used to

gain experience in how to analyze the data and to identify those
distributions
which can be measured accurately and calculated reliably.
We will break the program into four steps with each phase
contributing to a better understanding of QCD at hadron colliders.

In the first phase we use the PDF's obtained from global analyses
\cite{mrsd,cteq3,GRV94,mrsa}
and the associated $\al(M_Z)$ as input parameters.

Then by comparing data to theory we

can identify those cross sections which
are most sensitive to the input parameters.

Using run 1A data it has become clear
that certain distributions will be better than others in
determining the parton density functions and $\al$.
For example, the parton density functions can be constrained from
di-jet data using angular correlations \cite{UA1ang},
the same-side to opposite-side ratio \cite{CDFssos,GGKssos,MRSssos}
and via the triply differential cross section
\cite{Weerts,D0sign,GGK3D,ES3D,talks}
while the strong coupling can be determined from vector boson
production at large transverse momentum
\cite{UA2Wjets,UA1Wjets,D0Wjets}.

In the second phase we will
assume a given PDF set as being correct and extract
$\al$.  Measuring $\al$ at a hadron collider is rather
different than measuring $\al$ at LEP
with the most important difference being the fact that one can

measure $\al$ from momentum transfers as low as a few

GeV all the way up to 500 GeV
simultaneously and with high statistics \cite{talks}.
In this paper we will use the

one-jet inclusive transverse energy distribution

measured from the '88-'89 CDF data
with an integrated luminosity of 4.2 $\mbox{pb}^{-1}$

to illustrate this method.
The analysis can be repeated
for the current CDF and D\O\ data sets

increasing the integrated luminosity to well over
100 $\mbox{pb}^{-1}$. These increased statistics

will have a major impact
on  both the statistical and systematic error

relative to the '88-'89 data set.

The results in this paper are therefore just

an illustration of the method and no
detailed effort has been made to determine the

experimental systematic errors thoroughly. This would
require detailed knowledge of the

correlation matrix for the systematic error which
is not readily available.
While the value of $\al$ extracted in

this way cannot be considered on the same footing
as that measured at LEP
because the PDF's themselves are dependent on $\al$, this measurement
will nevertheless provide valuable information.

For example, the extracted $\al$ {\cal must} be consistent with
the $\al$ used in the PDF, or else the data is incompatible with

this particular set.
If one finds that the extracted $\al$ is
compatible with the PDF the measurement gives
an additional constraint on the PDF at large $x$ and $Q^2$.
Further, one can also study the
evolution of $\al$ for a wide range of momentum transfers.\footnote{
An alternative approach has been followed by UA2 \cite{UA2Wjets} and
is now
being pursued actively by both CDF and D\O\ .

Here, one uses PDF's
which are fitted to the Deep Inelastic Scattering (DIS) data set

for several values of $\al(M_Z)$. This
allows simultaneous variation of $\al$ in the PDF and matrix elements

leading to a consistent $\al(M_Z)$ extracted from

the combined DIS and hadronic data
set which can then be directly compared to the LEP value of $\al$.}

In the third phase we determine both

the PDF's and $\al$ simultaneously
using the triply differential di-jet inclusive distributions
\cite{CDFssos,Weerts,D0sign},
possibly including flavor tagging, yielding an $\al$

that is completely independent of the DIS data set.

In principle the measurement is very simple,

the parton fractions are determined
by summing over the rapidity weighted

transverse momenta of the particles

produced by the hard scattering,

$x_{1,2} = (\sum_i E_T^{i}e^{\pm \eta_i})/\sqrt{S}$
\cite{GGK3D,ES3D}.
However, since it is impossible to measure and identify all the

particles associated with the hard scattering, we are forced to rely
on

higher order calculations to estimate the unobserved

radiation. It might therefore be prudent to separate

this phase into two steps
by first determining the distribution of gluons in the

proton and assuming the
distribution of charged partons is determined by the DIS data set.
The reason for this is that in deep inelastic scattering
the virtual photon directly probes the
charged parton distributions.
The effects of the gluon distribution enter first

at next-to-leading order and cause, for example,

scaling violations in

the slope of $F_2$.
On the other hand, in hadron colliders,

the gluon density enters at lowest order

and a more direct measurement should be possible.
For example, by using the triply differential di-jet data one probes
the gluon distribution directly with essentially unlimited
statistics.
After the gluon distribution has been successfully extracted in this
manner,
one can include the triply differential $V +$ jet data

(where $V = W, Z,\gamma$) to extract the charged PDF's.
A succesful determination of both $\al$ and the

PDF's from the hadronic data set over a wide
range of momentum transfers would be an important

test of QCD and its consistent description using perturbative QCD.
The measured PDF's and $\al$ can directly be used in

other physics analyses at hadron colliders thereby
reducing the experimental systematic uncertainties considerably.

After completing the program outlined above,
one can then test QCD and

the gauge group responsible for the strong interactions
from first principles.  This would be the final phase
and is quite similar to the efforts at LEP \cite{LEPgauge}.
Of course, in hadron colliders there is the
additional interesting feature that the PDF

and its evolution are also predicted by the gauge group.
All together, this will give an accurate measurement
of the gauge nature of the strong interactions and quantify how well
the
data set fits the QCD theory.

This program is an achievable goal for the Main Injector run where,
because of the expected high luminosities and small

experimental errors, we expect to see deviations from
the next-to-leading order predictions, even without assuming new
physics.
This makes it crucial we understand the uncertainties

related to the PDF's and QCD very well. By
determining $\al$ and the PDF's within one experiment one

can identify which parts of the theoretical calculation
are important and try to improve them.
Furthermore, if significant deviations from
next-to-leading order show up,
it will be easier to identify possible problems in the

theory or conclude that the deviations are due to new physics.

An added bonus is that

the PDF's and $\al$ determined at large $x$ and $Q^2$ can

immediately be used in other physics analyses which

naturally occur at similar $x$ and $Q^2$ values,
thereby further reducing the systematic errors

associated with luminosity, $\al$, PDF's, etc.
Finally one can test QCD in a very rigorous manner

by comparing the parton density functions determined in
both deep inelastic scattering and

the hadron collider at a common scale.
Eventually, this might lead to a unified global fit of the PDF's to
{\it all} hadronic data.

In sec.~2 we will discuss the theoretical issues involved in
extracting $\al$ and will set up a general framework to extract $\al$
from a given data set. This framework is applied in sec.~3
to the one-jet inclusive transverse energy distribution.
Section 4 contains a brief description of the CDF data, while the
detailed results for the determination of $\al$ and it's evolution
are presented in sec.~5.
The conclusions summarize our main results and
briefly discuss
the prospects for measuring $\al$ at the TEVATRON.

\section{Theoretical considerations}

\subsection{Running coupling constants}
\begin{figure}[th]\vspace{10cm}
\includegraphics{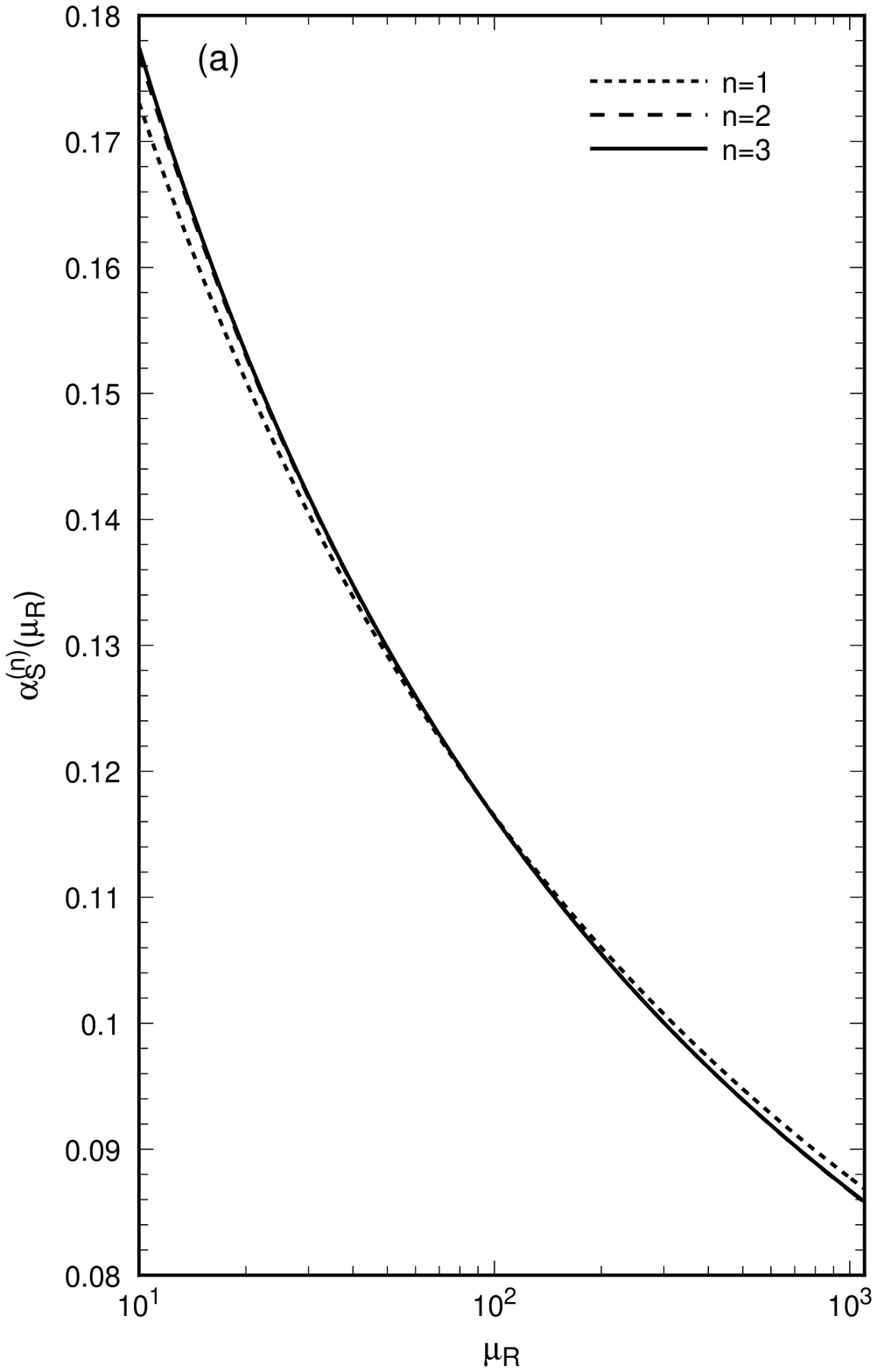}
\includegraphics{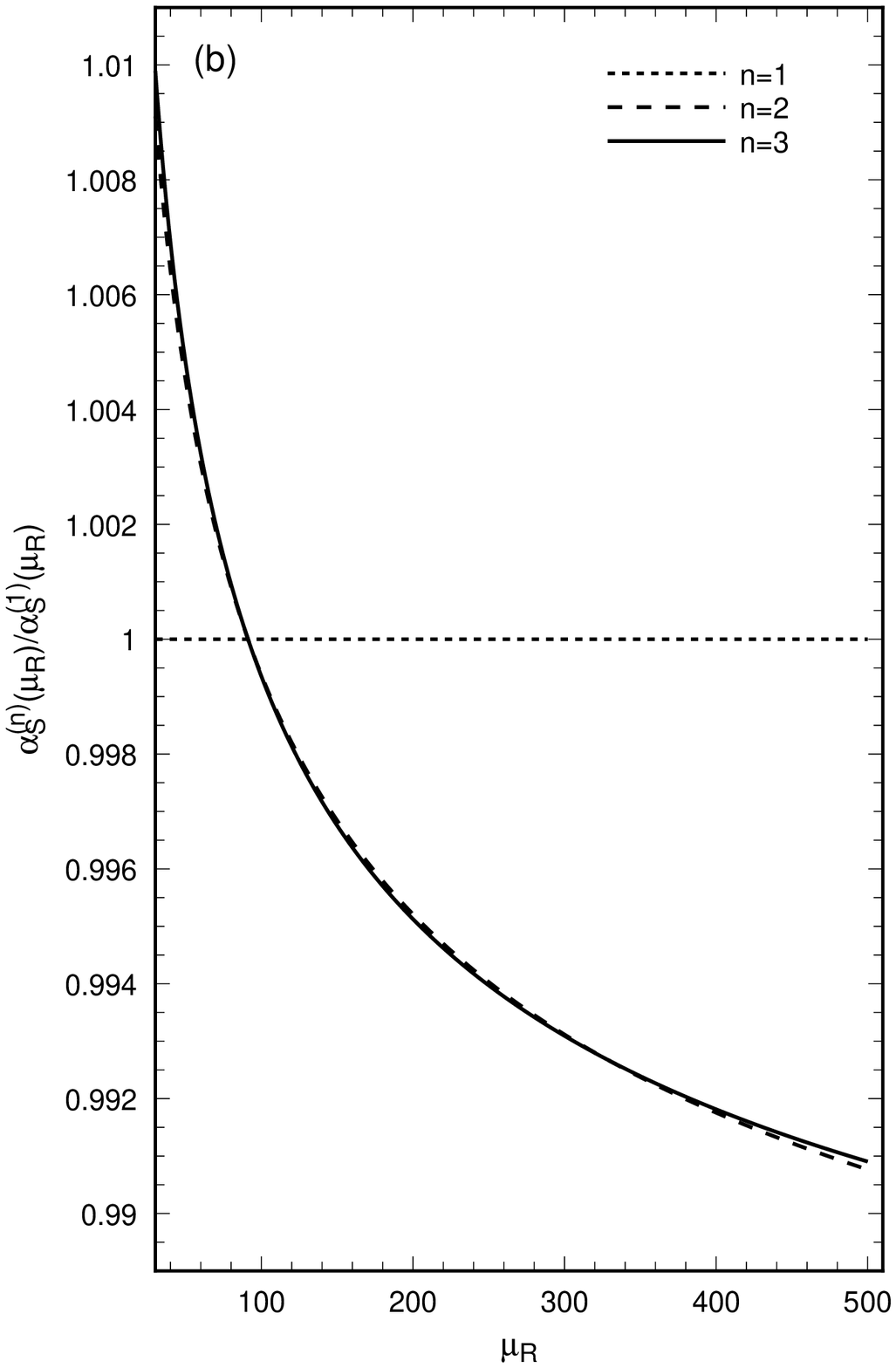}
\caption[]{Comparison between the 1st, 2nd and 3rd order running of
$\al(\mu_R)$
for $\al(M_Z)=0.118$.
Fig.~1a shows $\al^{(n)}(\mu_R)$ from 10 GeV to 1000 GeV while

Fig.~1b gives the relative change with respect to the 1st order
running
over the relevant energy range from 30 GeV to 500 GeV.}
\end{figure}

In order to calculate an observable

${\cal O}^{data}$ within perturbative QCD
we have to introduce the renormalization scale $\mu_R$.

However, no matter what scale we choose, it
cannot affect the prediction for the physical observable.
This statement can be formalized in the

renormalization group equation for the running coupling constant
$\al(\mu_R)$.
Both the coupling
constant and the matrix element coefficients

depend on the renormalization scale while
the physical quantity does not.
This means that

when measuring $\al$ we have to specify the

renormalization scale
so that the extracted $\al$ will
be the value of $\al$ at that particular renormalization
scale.
Of course, all possible choices of the
renormalization scale are related to each other
by virtue of the renormalization group equation.
Relative to a fixed scale, $M_Z$, the $n$-loop
running coupling constant at scale $\mu_R = \lambda M_Z$
is given by,
\begin{equation}\label{runningalpha}
\al(\mu_R)=\frac{\al(M_Z)}{1+\al(M_Z) L^{(n)}(\lambda)}
\end{equation}
where,
\begin{eqnarray}
L^{(1)}(\lambda)&=&b_0\log\left(\lambda\right),\\
L^{(2)}(\lambda)
&=&(b_0+b_1\al(M_Z))\log\left(\lambda\right),\label{2-loop}\\
L^{(3)}(\lambda)
&=&(b_0+b_1\al(M_Z)+b_2^{\overline{\mbox{\tiny
MS}}}\al^2(M_Z))\log\left(\lambda\right)
      -\frac{b_0 b_1}{2}\al^2(M_Z)\log^2\left(\lambda\right).
\label{3-loop}
\end{eqnarray}
The first three coefficients of the Callan-Symanzik $\beta$-function
are given by \cite{b0,b1,b2},
\begin{eqnarray}\label{bcoef}
b_0&=& \frac{11 N_c - 2 n_f}{6\pi}, \nn
b_1&=& \frac{34 N_c^2-13 N_c n_f +3 n_f/N_c}{24\pi^2},\\
b_2^{\overline{\mbox{\tiny MS}}}&=&
\frac{5714 N_c^3-3391 N_c^2 n_f
+ 224 N_c n_f^2 + 507 n_f + 54 n_f/N_c^2 -66 n_f^2/N_c}
{3456\pi^3}, \nonumber
\end{eqnarray}
where $N_c$ is the number of colors and $n_f$ the

number of active flavors.\footnote{The numerical values for the

$\beta$-function coefficients are:
$b_0=1.2202$, $b_1=0.4897$ and $b_2^{\overline{\mbox{\tiny
MS}}}=0.1913$

for $N_c=3$ and $n_f=5$.}
Note that while $b_0$ and $b_1$ are independent

of the renormalization scheme, $b_2$ is
renormalization scheme dependent.
The expression given here is for the  $\overline{\mbox{MS}}$
scheme which we use throughout the paper.

For the processes we will consider, the momentum transfer

ranges from 30 GeV up to around 500 GeV.

In Fig.~1a we show the running $\al$ in this range for

$\al(M_Z)=0.118$. We see that the differences between the
evolution at different orders is rather small in the relevant energy
range,
essentially because $\al(M_Z)$ and $\log(\lambda)$ are both small.
To see the differences more clearly,

Fig.~1b shows the relative change in $\al(\mu_R)$

with respect to the 1st order evolution.

The percentage change between
1st and 2nd order evolution
in the range from 30 GeV to 500 GeV is less than $\pm$1\%
and could be safely ignored with the present level of theoretical

and experimental accuracy.
In addition, the difference

between 2nd and 3rd order is completely negligible.
Therefore, in the rest of the paper we will use the 2-loop evolution
as given in

eq.~\ref{2-loop}.

\subsection{Extracting $\al$}

Measuring $\al$ from an observable ${\cal O}^{data}$ is, in
principle,

rather straightforward {\em provided} higher twist
effects and other non-perturbative effects are small so that the
perturbative expansion can be considered a reliable estimate of the
cross section.
In other words,
we calculate the perturbative expansion ${\cal O}^{pert}$
and equate it with the data,
\begin{equation}
{\cal O}^{data}\equiv {\cal O}^{pert}.
\end{equation}
For example, for the single jet inclusive transverse energy
distribution,
$$
{\cal O} = \frac{d\sigma}{dE_T},
$$
there is a good agreement between perturbative calculations and
the data over 7 orders of magnitude in the cross section

in the range $30\leq E_T\leq 500$ GeV \cite{EKS1,CDFpub}.

By comparing data with theory for each $E_T$-bin,
we make many independent measurements of

$\al$ at a specified

renormalization scale assuming no correlated bin-to-bin

experimental systematic errors.

The perturbative expansion can be written,
\begin{equation}\label{Born}
{\cal O}^{pert}
=
\al^m(\mu_R)\hat {\cal O}^{(0)} K^{(\infty)}(\al(\mu_R),\mu_R/Q_R),
\end{equation}
where the scale $Q_R$ is the characteristic scale for the observable
under

consideration which will be the transverse

energy of the jet for the single jet inclusive transverse energy
distribution.
The Born-prediction is given by $\al^m(\mu_R)\hat {\cal O}^{(0)}$
and all the higher order corrections
are contained in the $K$-factor,
\begin{equation}\label{Kfactor}
K^{(n)}(\al(\mu_R),\mu_R/Q_R)=1 +\sum_{l=1}^{n} \al^l(\mu_R)
k_l(\mu_R/Q_R).
\end{equation}
For the single jet inclusive transverse energy distribution,

$m = 2$ and the $K$-factor is currently

known up to next-to-leading order, giving $K^{(1)}$.

Once the $K$-factor has been calculated up to the $n$-th order in
$\al$,
the first step in extracting the $n$-th order $\al^{(n)}$

is to determine the leading order $\al^{(0)}$ together with
its experimental uncertainties.

This is simply given by the ratio
of the data over the leading order coefficient $\hat {\cal O}^{(0)}$
\begin{equation}\label{loalf}
{\al^{(0)}}=\sqrt[m]{\frac{{\cal O}^{data}}{\hat {\cal O}^{(0)}}},
\end{equation}
and does not depend on the renormalization

scale.\footnote{For hadronic collisions the Born term $\hat {\cal
O}^{(0)}$

will have an implicit dependence on the factorization scale, $\mu_F$.
Throughout, we specify $\mu_F = Q_R$.}

While the determination of the leading order $\al$

has no useful theoretical
interpretation it is nevertheless a very convenient

manner in which to parametrize the data.

In principle, all we need from an experiment
is the leading order $\al$ as given in eq.~\ref{loalf}
together with the experimental errors. From here

we can determine $\al^{(n)}$ values without
referring back to the original data.
For example, given  $K^{(n)}$, the
$n$-th order $\al$ is given by,
\begin{equation}\label{n-order}
{\al^{(n)}}(\mu_R)=
\frac{\al^{(0)}}{\sqrt[m]{K^{(n)}(\al^{(n)}(\mu_R),\mu_R/Q_R)}},
\end{equation}
so that $\al^{(n)}(\mu_R)$ are just the roots of
the $(m+n)$-th order polynomial,
\begin{equation}
[{\al^{(n)}}(\mu_R)]^m
\left(1 +\sum_{l=1}^{n} [\al^{(n)}(\mu_R)]^l k_l(\mu_R/Q_R)\right)
-[\al^{(0)}]^m=0.
\end{equation}

\subsection{Theoretical Uncertainty}

For the processes of interest, only the first order corrections

are currently known.
Therefore, we use the leading order $\al^{(0)}$ (with experimental
errors)

and solve the $(m+1)$th order polynomial,
\begin{equation}\label{alfpoly}
k_1(\mu_R/Q_R)
[\al^{(1)}(\mu_R)]^{(m+1)}+[\al^{(1)}(\mu_R)]^m-[\al^{(0)}]^m = 0,
\end{equation}
with $\mu_R = \mu_0$ to find $\al^{(1)}(\mu_0)$.
For the single jet inclusive transverse energy distribution,

we choose $\mu_0 = Q_R = E_T$.

To estimate the theoretical uncertainty we can extract $\al$
at renormalization scale $\mu_R=\lambda \mu_0$ which we subsequently
evolve back to scale $\mu_0$ using the 2-loop renormalization group
running of $\al$.
Quantitatively this means we solve eq. \ref{alfpoly} with

$\mu_R=\lambda \mu_0$ and determine,
\begin{equation}\label{evol1}
\al(\mu_0;\mu_R=\lambda \mu_0)=\frac{\al(\lambda \mu_0)}
{1+\al(\lambda \mu_0) L^{(2)}(1/\lambda)}.
\end{equation}
By defining,

\begin{equation}\label{terror}
\frac{\Delta\al(\lambda)}{\al}\equiv
\frac{\al(\mu_0;\mu_R=\lambda \mu_0)-\al(\mu_0;\mu_R=\mu_0)}
{\al(\mu_0;\mu_R=\mu_0)},
\end{equation}
we can estimate the theoretical uncertainty in $\al$.

The reason $\Delta\al$ is non-zero is due to the fact that the NLO
coefficient
$k_1(\mu_R=\lambda \mu_0)$ is only the first order
correction and rest of the higher order
corrections are neglected. The behavior of $k_1$ under
renormalization
scale changes is,
\begin{equation}\label{kevolv}
k_1(\mu_R=\lambda \mu_0)=k_1(\mu_R=\mu_0) + m b_0\log(\lambda)\ .
\end{equation}
This shift changes the solution of eq. \ref{alfpoly} giving us

$\al(\lambda \mu_0)$. However, evolving back to $\mu_R = \mu_0$ using

eq.~\ref{evol1} does not exactly

match the change due to eq.~\ref{kevolv}. In fact the shift in $\al$
due to
this mismatch

has a relatively simple form for small $\log(\lambda)$,
\begin{equation}\label{logtan}
\frac{\Delta\al(\lambda)}{\al}\sim\al^2
\left(
\frac{(m+1)b_0 k_1(\mu_R=\mu_0)}{ m+(m+1)\al k_1(\mu_R=\mu_0) }
+ b_1 \right)
\log(\lambda)
 + {\cal O}(\log^2(\lambda)).
\end{equation}

To extract the central value of $\al(\mu_0)$
and its theoretical uncertainty we can follow many different
procedures.
We will outline two of them here.

\begin{description}
\item{\underline {Method~I}}

The first procedure is rather straightforward. We take $\mu_R=\mu_0$
as
the central scale and vary the renormalization scale between
$\mu_R=\mu_0/2$
and $\mu_R=2 \mu_0$ to estimate the uncertainty. Explicitly this
means,
\begin{eqnarray}\label{methodI}
\al^{(1)}(\mu_0) &=&
\frac{1}{2}\Big (\al(\mu_0;\mu_R=2 \mu_0) +
\al(\mu_0;\mu_R=\mu_0/2)\Big ),\nn
\Delta\al^{(1)}(\mu_0) &=&
\frac{1}{2}\Big (\al(\mu_0;\mu_R=2 \mu_0) -
\al(\mu_0;\mu_R=\mu_0/2)\Big ).
\end{eqnarray}

\item{\underline {Method~II}}

The second procedure is based on the fact that $\Delta\al / \al$ has
a minimum.
This occurs when
$\lambda=\lambda_0=\exp(-k_1(\mu_R=\mu_0)/mb_0)$

so that the first order correction vanishes,
$k_1(\mu_R=\lambda_0 \mu_0) = 0$ and
$\al^{(0)}=\al^{(1)}(\mu_R=\lambda_0 \mu_0)$.

Now we
can define $\al^{(1)}(\mu_0)$ and the theoretical uncertainty
in the following manner,
\begin{eqnarray}\label{methodII}
\al^{(1)}(\mu_0)&=&\frac{1}{2}\Big(\al(\mu_0;\mu_R=\mu_0)
+\al(\mu_0;\mu_R=\lambda_0 \mu_0)\Big), \nn
\Delta\al^{(1)}(\mu_0)&=&\frac{1}{2}\Big(\al(\mu_0;\mu_R=\mu_0)
-\al(\mu_0;\mu_R=\lambda_0 \mu_0)\Big).
\end{eqnarray}

\end{description}

There are two major differences between the two methods of

estimating the theoretical
uncertainty.
First,
the estimated theoretical uncertainty is
generally larger in method I  than in method II
and second,
the central value of $\al(\mu_0)$ using method II is slightly lower,
but by construction it lies within the range of uncertainty of method
I.

\section{The one-jet inclusive transverse energy distribution}

The one-jet inclusive transverse energy distribution ($d\sigma/d
E_T$)
has a straightforward interpretation: the transverse
energy of the leading jet ($E_T$) is directly related to the
impact parameter $b_{impact}$ (or the distance scale) in the
underlying hard
parton-parton scattering by the relation
\begin{equation}
\left(\frac{b_{impact}}{1\ {\rm fm}}\right)
=0.0507\times\left(\frac{E_T}{100\ {\rm GeV}}\right)^{-1}.
\end{equation}
Therefore by studying this particular distribution we probe rather
directly the physics over a wide range of distance scales within one
single measurement. For the published CDF data,
the transverse energies range from 30 GeV up to 500 GeV. In other
words, we probe the dynamics of the parton-parton
scattering from a distance scale of 0.169~fm all the way down to
0.01~fm.
The obvious quantity to study is therefore
$\al$ extracted at renormalization scale $\mu_R = E_T$.
Subsequently we can test QCD

by comparing the measured $\al$ at the different distance scales with
the running $\al$ predicted by QCD.
The comparison will be sensitive to new physics, the most

obvious being substructure of the quarks. However,

deviations from QCD at small distance scales will also
show up as violations of the running of $\al$.
To perform the comparison with QCD we will use two methods, each of
which has its own interest. The first one assumes the evolution is
correctly

given by QCD to extract $\al$ at a common scale, $\mu_R = M_Z$,
while the second method quantifies  deviations from purely QCD-like
 evolution.\\

\subsection{The QCD-fit}

In the first method we {\it assume} the correctness of QCD to
describe
the parton-parton scattering at all distance scales relevant in this
measurement. This enables us to extract the best possible value
of $\al(M_Z)$ using a given data set.
Each $E_T$-bin in the differential cross section
gives an independent measurement of $\al(E_T)$ which
we subsequently can evolve to $\al(M_Z)$ using the
renormalization group equation.
The published CDF data set has 38 individual $E_T$-bins,
and therefore yields 38 independent measurements of $\al(M_Z)$, so
that the statistical
error will be negligible compared to the common systematic error
which
has two components.
The first is the calorimeter response
correction together with fragmentation/hadronization effects and

the second is the luminosity uncertainty.

The luminosity uncertainty can be reduced

using the $W$-boson production cross section ($\sigma_W$)
as a luminosity measurement. Experimentally this simply involves
counting the number of $W$-boson events and normalizing $d\sigma/d
E_T$

respectively, i.e. we study $1/\sigma_W d\sigma/d E_T$
so that the luminosity uncertainty cancels.
Theoretically, $\sigma_W$ is the best known cross section at
hadron colliders, known up to 2-loop QCD corrections \cite{Neerven}.

Therefore

$1/\sigma_W d\sigma/d E_T$ can be calculated

consistently order by order in perturbative QCD and compared
to experiment to extract $\al$.

This method of normalizing the cross section

can easily be generalized to all observables.

\subsection{The Best-fit}

It may turn out that the measured $\al(M_Z)$ is not independent

of the $E_T$ values it was extracted from.
This indicates either deficiencies in the input
PDF or, more interestingly, deviations from the underlying QCD
theory.
Parametrizing possible deviations from the QCD condition
$\partial\al(M_Z;\mu_R=E_T)/\partial E_T = 0$ will give us an
excellent check on the theory.
We therefore quantify deviations
from QCD by allowing

$\partial\al(M_Z;\mu_R=E_T)/\partial E_T = f(E_T) \neq 0$.

The size of the deviation $f(E_T)$ and its uncertainty will tell us
how well the data set fits QCD. More interestingly, by evolving
the fit to $\al(M_Z;\mu_R=E_T)$ back to $\al(E_T)$ we obtain

the ``Best-fit''
prediction for  the evolution of  $\al$. By extrapolating the fit to
larger and smaller scales we find the permissible range
of evolution for $\al(\mu_R)$ {\it allowing} for
small deviations from QCD in the current data set.

We can then compare with $\al$ measurements
at different energy scales and see how compatible the deviations
are with the other measurements, in particular the slower running
that may be suggested by the low energy data \cite{slower}.
While the systematic error dominates in the ``QCD-fit'' method, here
the systematic error (including the theoretical uncertainty) will
merely
affect the overall normalization  of the $\al$ evolution and not the
shape.

In fact by normalizing the curve to the world average value of
$\al(M_Z)$ we can completely remove the systematic error.

\section{The CDF data}

The CDF data used in this analysis is from the '88-'89 TEVATRON
collider
run at Fermilab which yielded an integrated luminosity of 4.2
pb$^{-1}$.
The data was taken from the preprint version \cite{CDFpreprint} of
the published
letter \cite{CDFpub}. The preprint tabulates the results together
with the
separated statistical and systematic errors.
Unfortunately the error-analysis in the current paper is limited
by the fact that the published results do not have the necessary
detailed discussion of the systematic errors needed for a  more
rigorous error treatment. We will use an ad-hoc procedure to
separate a common systematic error and a bin-by-bin statistical
error. Also the removal of the luminosity error using the $W$-boson
cross section cannot be applied, as this would require a careful
simultaneous study of the $W$-boson and jet data. However, both the
CDF
and D\O\ collaborations can incorporate a proper error analysis
and removal of the luminosity error using the new run 1A/B data sets.

Our purpose here is to illustrate the methods rather than

produce a definitive measurement and error analysis.

The one-jet inclusive transverse energy

distribution of CDF is constructed by including
{\it all} the jets, which are defined according to the

Snowmass algorithm with a conesize of 0.7 \cite{Snowmass}, in the
pseudo-rapidity range between 0.1 and 0.7. This means

that this particular
distribution is not exactly the transverse energy distribution of the
leading jet as would be preferred for the $\al$ measurement, but
contains some

softer jets. However, although small deviations
from the leading jet $E_T$ distribution can be expected at small
$E_T$,
for high-$E_T$ bins there is virtually no difference between the two
distributions.

Some of the features of the particular data taking for the '88-'89
run
will be reflected in the $\al$ measurement. First of all, the
statistical
error is affected by the different event triggers, each with its own

prescaling factor. The off-line $E_T$ cuts (ensuring 98\% efficiency
in
the data taking) for the three triggers are 35 GeV (with prescaling
factor

1 per 300 events), 60 GeV (1 per 30 events) and 100 GeV (not
prescaled) \cite{CDFpub}.
This means the statistical errors fall into three distinct regions
which
eventually will be reflected in the statistical error on $\al$.

Second, the systematic error is due to the luminosity measurement
and the detector response combined with
the hadron distribution within the jets (which is modelled by

fragmentation/hadronization Monte Carlo's). The systematic error
quoted
contains all these uncertainties and most of the error will
be common to all the bins.
Apart from the luminosity error,

the systematic error falls into two separate regions.

Below $E_T = 80$~GeV, the systematic error is large,
decreasing from as high as $\pm$60\% at 35~GeV to $\pm$22\%
at 80~GeV. As a result, the $\al$ measurement below 80~GeV is
strongly affected by short range correlated systematic errors.

For $E_T > 80$~GeV, however, the systematic error is fairly constant
with a typical value of $\pm$22\%, making the short range systematic
error correlations small. Again these characteristics of the data
will eventually be reflected in the measured $\al$.

\section{Determining $\al$}

To extract $\al$ we use the next-to-leading order parton level
Monte Carlo JETRAD \cite{jetrad}
which is based on the techniques described in refs.~\cite{GG,dyrad}
and the matrix elements of ref.~\cite{ES}.

The cuts
and jet algorithm applied directly to the partons,
were modelled as closely as possible to the experimental
set-up. Using the Monte Carlo we calculated the Born coefficient

${\hat O}^{(0)}$ and the next-to-leading order coefficient $k^{(1)}$
as defined
in eqs. \ref{Born} and \ref{Kfactor} for the
MRSA$^\prime$ PDF set of ref.~\cite{mrsa}.

These distribution functions use the low-$x$ $F_2$ data from the 1993
data taking run at HERA.
However, we are mainly concerned with $x$ values typically greater

than few $\times 10^{-2}$, and there is little impact from HERA

data in this range.
To see this, we also consider the older MRSD0$^\prime$

and MRSD-$^\prime$ parameterisations \cite{mrsd}.
Using eq.~\ref{loalf} we determine the leading order $\al^{(0)}$ from
the

CDF data including the statistical and systematic errors. The results
are listed in Table~1 together with the next-to-leading order
coefficients
determined at a renormalization/factorization scale equal to the
$E_T$ value of
the bin and the Monte Carlo integration error.

This table contains all the information needed to extract the

next-to-leading order $\al$.

\subsection{Measurement of $\al^{(1)}(E_T)$}
\begin{table}[p]\begin{center}
\vspace{-1.5cm}
\begin{tabular}{|c|c|c|c|c|}\hline
$E_T$ & \multicolumn{4}{c|}{MRSA$^\prime$} \\ \cline{2-5}
(GeV) & $\al^{(0)}$ & $k_1$ & $\al^{(1)}(E_T)$ & $\al^{(1)}(M_Z)\pm
0.008$ \\ \hline
 35.48 & ${0.163 \pm 0.003}^{+.027}_{-.049}$ & $2.90 \pm 0.24$ &

 $0.138 \pm 0.028 \pm 0.007$ & $0.118 \pm 0.013 \pm 0.005$ \\
 41.63 & ${0.154 \pm 0.001}^{+.025}_{-.044}$ & $2.94 \pm 0.26$ &

 $0.131 \pm 0.026 \pm 0.006$ & $0.116 \pm 0.013 \pm 0.005$ \\
 47.61 & ${0.149 \pm 0.001}^{+.024}_{-.039}$ & $3.18 \pm 0.29$ &

 $0.126 \pm 0.023 \pm 0.006$ & $0.114 \pm 0.011 \pm 0.005$ \\
 53.54 & ${0.149 \pm 0.002}^{+.022}_{-.033}$ & $3.42 \pm 0.27$ &

 $0.125 \pm 0.020 \pm 0.006$ & $0.115 \pm 0.009 \pm 0.005$ \\
 59.93 & ${0.143 \pm 0.003}^{+.021}_{-.029}$ & $3.04 \pm 0.24$ &

 $0.122 \pm 0.019 \pm 0.005$ & $0.115 \pm 0.009 \pm 0.004$ \\
 66.23 & ${0.144 \pm 0.003}^{+.019}_{-.024}$ & $3.34 \pm 0.23$ &

 $0.122 \pm 0.016 \pm 0.005$ & $0.116 \pm 0.007 \pm 0.005$ \\
 72.29 & ${0.145 \pm 0.001}^{+.019}_{-.021}$ & $2.80 \pm 0.25$ &

 $0.125 \pm 0.015 \pm 0.005$ & $0.121 \pm 0.006 \pm 0.005$ \\
 78.15 & ${0.147 \pm 0.002}^{+.018}_{-.018}$ & $3.11 \pm 0.24$ &

 $0.125 \pm 0.013 \pm 0.005$ & $0.122 \pm 0.005 \pm 0.005$ \\
 83.81 & ${0.153 \pm 0.002}^{+.018}_{-.016}$ & $3.57 \pm 0.21$ &

 $0.127 \pm 0.012 \pm 0.006$ & $0.125 \pm 0.004 \pm 0.006$ \\
 89.31 & ${0.148 \pm 0.002}^{+.018}_{-.016}$ & $3.35 \pm 0.25$ &

 $0.125 \pm 0.013 \pm 0.006$ & $0.124 \pm 0.005 \pm 0.006$ \\
 94.82 & ${0.147 \pm 0.003}^{+.018}_{-.016}$ & $3.39 \pm 0.22$ &

 $0.123 \pm 0.013 \pm 0.006$ & $0.124 \pm 0.005 \pm 0.006$ \\
100.19 & ${0.145 \pm 0.003}^{+.018}_{-.016}$ & $3.73 \pm 0.23$ &

 $0.120 \pm 0.012 \pm 0.005$ & $0.122 \pm 0.005 \pm 0.006$ \\
105.60 & ${0.144 \pm 0.004}^{+.017}_{-.016}$ & $3.45 \pm 0.23$ &

 $0.121 \pm 0.012 \pm 0.005$ & $0.124 \pm 0.005 \pm 0.006$ \\
111.04 & ${0.143 \pm 0.004}^{+.017}_{-.015}$ & $3.28 \pm 0.23$ &

 $0.121 \pm 0.012 \pm 0.005$ & $0.125 \pm 0.005 \pm 0.006$ \\
116.44 & ${0.135 \pm 0.001}^{+.017}_{-.016}$ & $3.77 \pm 0.22$ &

 $0.113 \pm 0.012 \pm 0.005$ & $0.118 \pm 0.005 \pm 0.005$ \\
121.76 & ${0.136 \pm 0.001}^{+.017}_{-.016}$ & $3.52 \pm 0.22$ &

 $0.115 \pm 0.012 \pm 0.005$ & $0.120 \pm 0.005 \pm 0.005$ \\
127.12 & ${0.135 \pm 0.001}^{+.017}_{-.016}$ & $3.50 \pm 0.21$ &

 $0.114 \pm 0.012 \pm 0.005$ & $0.120 \pm 0.006 \pm 0.005$ \\
132.49 & ${0.134 \pm 0.001}^{+.017}_{-.016}$ & $3.77 \pm 0.23$ &

 $0.113 \pm 0.012 \pm 0.005$ & $0.119 \pm 0.005 \pm 0.005$ \\
137.77 & ${0.135 \pm 0.002}^{+.017}_{-.016}$ & $3.44 \pm 0.20$ &

 $0.114 \pm 0.012 \pm 0.005$ & $0.121 \pm 0.006 \pm 0.005$ \\
143.05 & ${0.134 \pm 0.002}^{+.017}_{-.016}$ & $3.77 \pm 0.22$ &

 $0.112 \pm 0.012 \pm 0.005$ & $0.120 \pm 0.006 \pm 0.005$ \\
148.48 & ${0.134 \pm 0.002}^{+.016}_{-.015}$ & $4.10 \pm 0.22$ &

 $0.111 \pm 0.011 \pm 0.005$ & $0.119 \pm 0.005 \pm 0.005$ \\
153.71 & ${0.132 \pm 0.002}^{+.016}_{-.015}$ & $3.78 \pm 0.22$ &

 $0.111 \pm 0.012 \pm 0.004$ & $0.120 \pm 0.006 \pm 0.005$ \\
158.93 & ${0.133 \pm 0.003}^{+.016}_{-.015}$ & $3.54 \pm 0.20$ &

 $0.112 \pm 0.011 \pm 0.004$ & $0.122 \pm 0.006 \pm 0.005$ \\
164.25 & ${0.133 \pm 0.003}^{+.016}_{-.015}$ & $4.26 \pm 0.19$ &

 $0.110 \pm 0.011 \pm 0.005$ & $0.119 \pm 0.005 \pm 0.006$ \\
171.89 & ${0.135 \pm 0.003}^{+.016}_{-.015}$ & $3.95 \pm 0.16$ &

 $0.113 \pm 0.011 \pm 0.005$ & $0.124 \pm 0.005 \pm 0.006$ \\
182.20 & ${0.135 \pm 0.003}^{+.015}_{-.014}$ & $3.98 \pm 0.14$ &

 $0.112 \pm 0.011 \pm 0.005$ & $0.125 \pm 0.005 \pm 0.006$ \\
193.04 & ${0.128 \pm 0.004}^{+.016}_{-.015}$ & $4.05 \pm 0.14$ &

 $0.107 \pm 0.011 \pm 0.004$ & $0.119 \pm 0.006 \pm 0.005$ \\
203.47 & ${0.116 \pm 0.005}^{+.017}_{-.016}$ & $3.75 \pm 0.14$ &

 $0.099 \pm 0.013 \pm 0.003$ & $0.111 \pm 0.008 \pm 0.004$ \\
215.73 & ${0.120 \pm 0.005}^{+.016}_{-.015}$ & $4.15 \pm 0.12$ &

 $0.101 \pm 0.012 \pm 0.004$ & $0.113 \pm 0.007 \pm 0.005$ \\
231.88 & ${0.122 \pm 0.006}^{+.015}_{-.014}$ & $4.22 \pm 0.11$ &

 $0.102 \pm 0.011 \pm 0.004$ & $0.116 \pm 0.007 \pm 0.005$ \\
246.86 & ${0.121 \pm 0.008}^{+.016}_{-.015}$ & $4.32 \pm 0.11$ &

 $0.101 \pm 0.012 \pm 0.004$ & $0.116 \pm 0.008 \pm 0.005$ \\
264.86 & ${0.115 \pm 0.011}^{+.016}_{-.015}$ & $4.17 \pm 0.10$ &

 $0.097 \pm 0.014 \pm 0.003$ & $0.112 \pm 0.010 \pm 0.004$ \\
281.96 & ${0.140 \pm 0.013}^{+.014}_{-.013}$ & $4.31 \pm 0.10$ &

 $0.114 \pm 0.013 \pm 0.005$ & $0.137 \pm 0.011 \pm 0.008$ \\
302.22 & ${0.137 \pm 0.022}^{+.014}_{-.013}$ & $4.28 \pm 0.09$ &

 $0.112 \pm 0.019 \pm 0.005$ & $0.136 \pm 0.019 \pm 0.007$ \\
322.87 & ${0.124 \pm 0.035}^{+.016}_{-.015}$ & $4.60 \pm 0.09$ &

 $0.103 \pm 0.028 \pm 0.004$ & $0.123 \pm 0.032 \pm 0.006$ \\
343.88 & ${0.154 \pm 0.056}^{+.016}_{-.014}$ & $4.55 \pm 0.08$ &

 $0.123 \pm 0.041 \pm 0.006$ & $0.155 \pm 0.056 \pm 0.010$ \\
380.72 & ${0.161 \pm 0.101}^{+.016}_{-.014}$ & $4.66 \pm 0.07$ &

 $0.127 \pm 0.078 \pm 0.007$ & $0.166 \pm 0.109 \pm 0.012$ \\
418.55 & ${0.188 \pm 0.143}^{+.018}_{-.017}$ & $4.86 \pm 0.07$ &
 $0.144 \pm 0.224 \pm 0.010$ & $0.201 \pm 0.251 \pm 0.020$ \\
\hline
\end{tabular}\end{center}
\caption[]{The extracted leading order $\al^{(0)}$ with statistical
and systematical errors based on the published CDF data
\cite{CDFpub}.
The higher order coefficient
$k_1$ for $\mu_R=\mu_F=E_T$ is also shown with its associated
MC-integration error.
The last two columns give $\al^{(1)}$ based on

the solution of eq.~\ref{alfpoly} with the combined
statistical/systematic errors from \cite{CDFpub} and
the theoretical uncertainty estimate using method I

as described in the text.}
\end{table}

\begin{figure}[t]\vspace{9cm}
\includegraphics{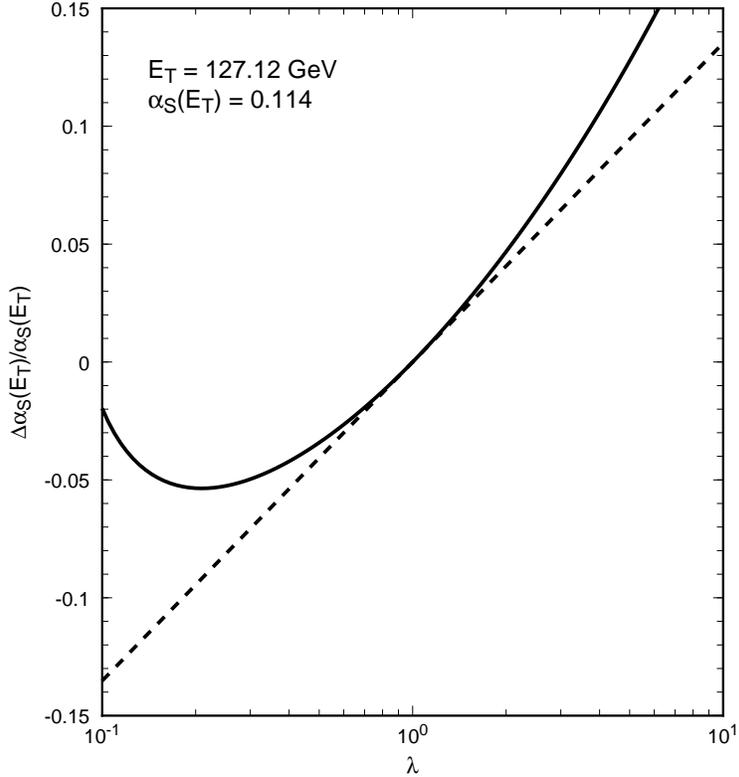}
\caption[]{The uncertainty in $\al(E_T)$ due to variation in the
renormalization scale (solid line).
Also shown is the logarithmic tangent of eq.~\ref{logtan} (dashed
line)
which has a simple analytic form.}
\end{figure}

Finally we are in a position to determine the next-to-leading order
$\al^{(1)}$
and the associated theoretical uncertainty.
We combine the experimental statistical and systematic errors on
the leading order $\al^{(0)}$ in quadrature and solve
eq.~\ref{alfpoly}
with $m=2$ and $\mu_0 = E_T$ to extract $\al^{(1)}(E_T)$.

In fig.~2 we show both the exact $\Delta\al(\lambda)/\al$ defined by

eq.~\ref{terror} and its
logarithmic tangent as given by eq.~\ref{logtan} for one $E_T$-bin.
We see that for $0.5 < \lambda < 2$, the linear approximation is
reasonable.
The extracted values of $\al^{(1)}(E_T)$

(with the associated theoretical errors)

defined by eqs.~\ref{methodI} and \ref{methodII}

for the 38 transverse energy bins
are given in Table~1.
Numerically,
the major difference between the two methods of estimating the
theoretical
uncertainty is that in method~I the estimated theoretical uncertainty
is
of the order of 4\%, while method~II gives a smaller theoretical
uncertainty
of typically 2\%.

The central value of $\al^{(1)}(E_T)$ using method~II is lower by
2\%,
but remains within the uncertainty of method~I.
For the rest of the paper we will use the

$\al(E_T)$ extraction based on method~I.

As a rough guide, changing to method~II means

reducing the theoretical uncertainty and
lowering the central value by 2\%.

\subsection{Measurement of $\al^{(1)}(M_Z)$}

The next step is to determine the strong coupling constant at $M_Z$
by
evolving from $\mu_R = E_T$ to $\mu_R=M_Z$

using the 2-loop evolution equation eq.~\ref{2-loop}.
To extract the common long range systematic error,

$\Delta\al^{sys}$, we employed the
following ad hoc procedure.

Because $\al(M_Z;\mu_R=E_T)$ is supposed to be independent
of $E_T$ we can define  $\Delta\al^{sys}$ such that,
\begin{equation}
\chi^2=\frac{1}{N_{bins}}\sum_{i=1}^{N_{bins}}
\frac{\left[\al^{(1)}(M_Z;\mu_R=E_T^{(i)})-<\al(M_Z)>\right]^2}
{\left[\Delta\al^{exp}(M_Z;\mu_R=E_T^{(i)})-\Delta\al^{sys}\right]^2}
= 1.
\end{equation}
Here $E_T^{(i)}$ refers to the specific bin-values.
This procedure gives us a value for $\Delta\al^{sys} = 0.008$, which

is then common to all values of $\al$. The remaining errors,

$\Delta\al^{stat}=\Delta\al^{exp}-\Delta\al^{sys}$,
are a combination of statistical errors and shorter range correlated
systematic errors. Fig.~3 and Table~1 display the values of
$\al(M_Z)$ extracted from the 38 $E_T$ bins with the associated

experimental statistical error and the estimate of the theoretical
uncertainty.
The systematic error is an overall factor of $\pm 0.008$.
We see that measured value of
$\al(M_Z)$ is essentially independent of $E_T$ for the MRSA$^\prime$
parton density functions. This is also true of the
MRSD0$^\prime$ and MRSD-$^\prime$ parameterisations.

\begin{figure}\vspace{9cm}
\includegraphics{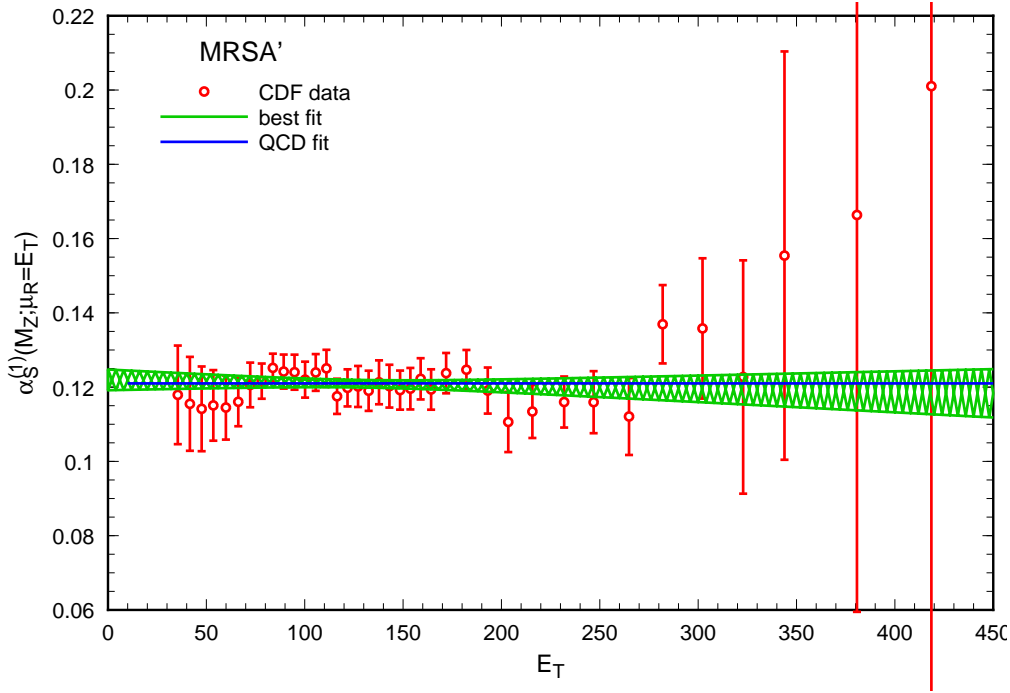}
\caption[]{The extracted $\al^{(1)}(M_Z, \mu_R=E_T)$ as a function
of $E_T$ for the MRSA$^\prime$ parameterisation.
The QCD-fit yields

$\al^{(1)}(M_Z)= 0.121\pm 0.001\pm 0.008\pm 0.005$.
The 68\% confidence level Best-fits are shown as shaded bands.}
\end{figure}

\subsection{The QCD-fit to $\al^{(1)}(M_Z)$}

To compare the obtained result with QCD we first assume that
next-to-leading order
QCD is sufficient to describe the data. In this case

$\partial\al(M_Z;\mu_R=E_T)/\partial E_T = 0$ and we can

perform an error weighted average
to obtain the average $\al^{(1)}(M_Z)$,
\begin{equation}
\al^{(1)}(M_Z)=\frac{1}{w}\sum_{i=1}^{N_{bins}} w_i
\al^{(1)}(M_Z;\mu_R=E_T^{(i)}),
\end{equation}
where,
\begin{eqnarray}
\frac{1}{w_i}&=& \Delta\al^{stat}(M_Z;\mu_R=E_T^{(i)}), \nn
w&=&\sum_{i=1}^{N_{bins}} w_i.
\end{eqnarray}
The resulting values for $\al^{(1)}(M_Z)$ are,
\begin{eqnarray}\label{meth1res}
\al^{(1)}(M_Z)&=& 0.119\pm 0.001\pm 0.008\pm 0.005\ \mbox{for
MRSD0$^\prime$} \nn
\al^{(1)}(M_Z)&=& 0.121\pm 0.001\pm 0.008\pm 0.005\ \mbox{for
MRSA$^\prime$}\\
\al^{(1)}(M_Z)&=& 0.124\pm 0.001\pm 0.008\pm 0.005\ \mbox{for
MRSD-$^\prime$} \nonumber
\end{eqnarray}
where the first error is the statistical error, the second error the
systematic error and the third error the theoretical uncertainty
estimate based
on method I\footnote{Using method II would lower $\al(M_Z)$ by 0.003
and reduce
the theoretical uncertainty to 0.003}.
We see that the error from using different PDF as input is
approximately
$\pm 0.002$.

A note of caution is in order.
In the analysis presented here, we have taken PDF's which have a
$Q^2$ evolution based on $\al^{DIS}(M_Z)=0.113\pm 0.005$ \cite{mrsa},
while the
explicit $\al$ in the matrix elements was varied. This can only
be consistent
if the extracted value of $\al(M_Z)$ is in agreement with
$\al^{DIS}(M_Z)$.
However, we see that this is indeed the case (c.f. \ref{meth1res})
once the

statistical, systematic and theoretical errors are combined.
With the new high statistic data sets
of run 1A/B, the statistical and systematic errors

will be significantly reduced and it will be necessary to utilise
PDF's with $Q^2$ evolution for a variety of $\al(M_Z)$ values.
Recently such PDF sets have come available \cite{GRVvar,mrsavar} so
that a
more consistent determination of $\al$ will be possible

once the new data becomes available.

\subsection{The Best-fit to $\al^{(1)}(M_Z)$}
\begin{table}[t]\begin{center}
\vspace{1cm}\label{chitable}
\begin{tabular}{|l|c|c|c|c|}\hline
PDF & $a$ & $b$ & $c$ & $d$ \\ \hline
MRSD0$^\prime$ & 0.119 &  0.0038 & 0.0008 & 11.0 \\
MRSA$^\prime$  & 0.120 & -0.0010 & 0.0010 & 6.8 \\
MRSD-$^\prime$ & 0.124 & -0.0014 & 0.0011 & 6.2 \\
\hline
\end{tabular}\end{center}
\caption{Best-fit results from minimum $\chi^2$-fit}
\end{table}
\begin{figure}\vspace{9cm}
\includegraphics{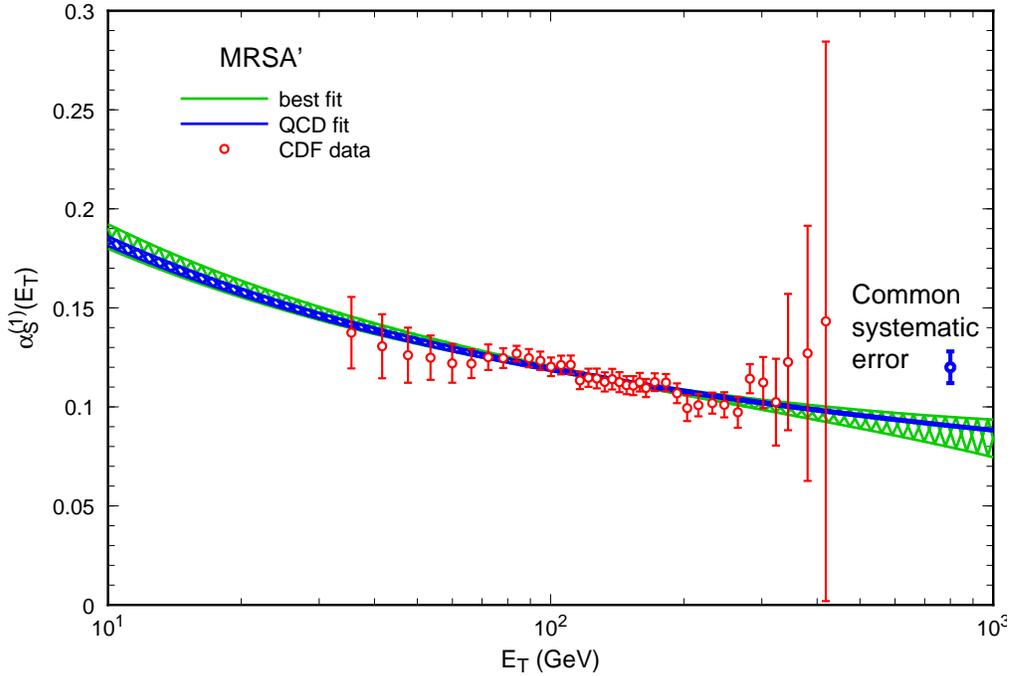}
\caption[]{The values of $\al^{(1)}(E_T)$

extracted from the published CDF data as a function
of $E_T$ together with $\al^{(1)}(E_T)$ from the QCD- and

Best-fits evolved from $M_Z$ to $E_T$ for

the MRSA$^\prime$ parameterisation.}
\end{figure}
\begin{figure}[th]\vspace{11cm}
\includegraphics{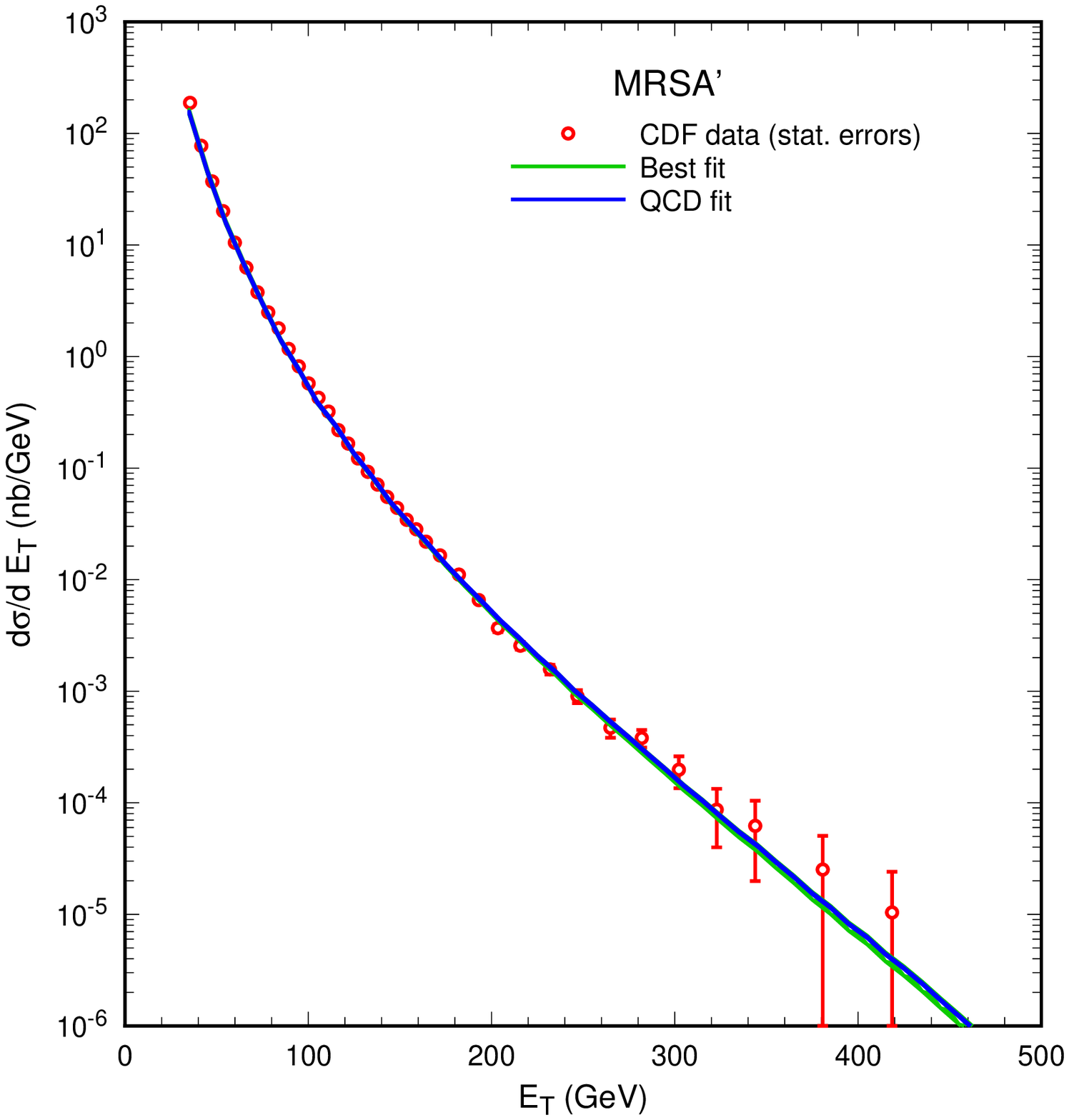}
\caption[]{Comparison of the one-jet inclusive transverse energy
distribution
evaluated using $\al^{(1)}(E_T)$ from the QCD- and

Best-fits evolved from $M_Z$ to $E_T$ for

the MRSA$^\prime$ parameterisation.}
\end{figure}

The second comparison with QCD we can perform is a check on the
running behavior of $\al$.
For such a check, the overall
systematic error is not important and the experimental error is
reduced
considerably.
This tests whether $\al(M_Z)$ is
independent of the distance scale at which the scattering takes
place.
To do this we no longer assume $\partial\al(M_Z;\mu_R=E_T)/\partial
E_T = 0$
but allow it to be a constant. If QCD is correct, the constant should
be zero
within errors.

The results from a minimal $\chi^2$ fit to a linear function in
$E_T$,

\begin{eqnarray}\label{minchi}
\al^{(1)}(M_Z;\mu_R=E_T) &=& a+
b\times\left(\frac{E_T}{E_T^0}-1\right)\nn
\Delta\al^{stat}(M_Z;\mu_R=E_T) &=&
c\times\sqrt{1+d\times\left(\frac{E_T}{E_T^0}-1\right)^2}\ ,
\end{eqnarray}
are given in Table~2 for the MRSD0$^\prime$, MRSA$^\prime$

and MRSD-$^\prime$ parameterisations.
The scale $E_T^0\simeq 130$ GeV gives the minimal one-sigma error.
The
common systematic error, $\Delta\al^{sys}=0.008$

is not affected by the fits.
The linear minimal-$\chi^2$ fits give a
perfect fit to QCD (i.e. no $E_T$ dependence) within one sigma over a
range from 30 GeV to 500 GeV for MRSA$^\prime$ and MRSD-$^\prime$,

while the MRSD0$^\prime$ results show a small but
insignificant dependence on the transverse energy which possibly
indicate
some problems with the underlying PDF set. It should be stressed that
these
results are highly non trivial and demonstrate the correctness of QCD
over
a wide range of momentum transfers (or distance scales) not
previously probed.

Although the statistics are rather poor at high $E_T$,  the new CDF
and
D\O\ results should give better results in the region above 200 GeV.

Next we can evolve the Best-fit result from $\al(M_Z;\mu_R=E_T)$
back to $\al(E_T)$ and extrapolate to
smaller and larger $E_T$ values to obtain the

measured running $\al$ and then compare that with the QCD prediction
from the QCD-fit.
This comparison is shown in fig.~4 where
we can see that
the measured evolution agrees perfectly with
the QCD evolution for the  MRSA$^\prime$  parameterisation.

On the other hand, if we use the Best-fit for the MRSD0$^\prime$ set,

we find a slower running of the coupling constant

which agrees very well with, in particular, the
low energy $\al$ measurements \cite{slower}.

The results from the new collider run will clarify this and test
the running $\al$ behavior much better.

\begin{figure}\vspace{9cm}
\includegraphics{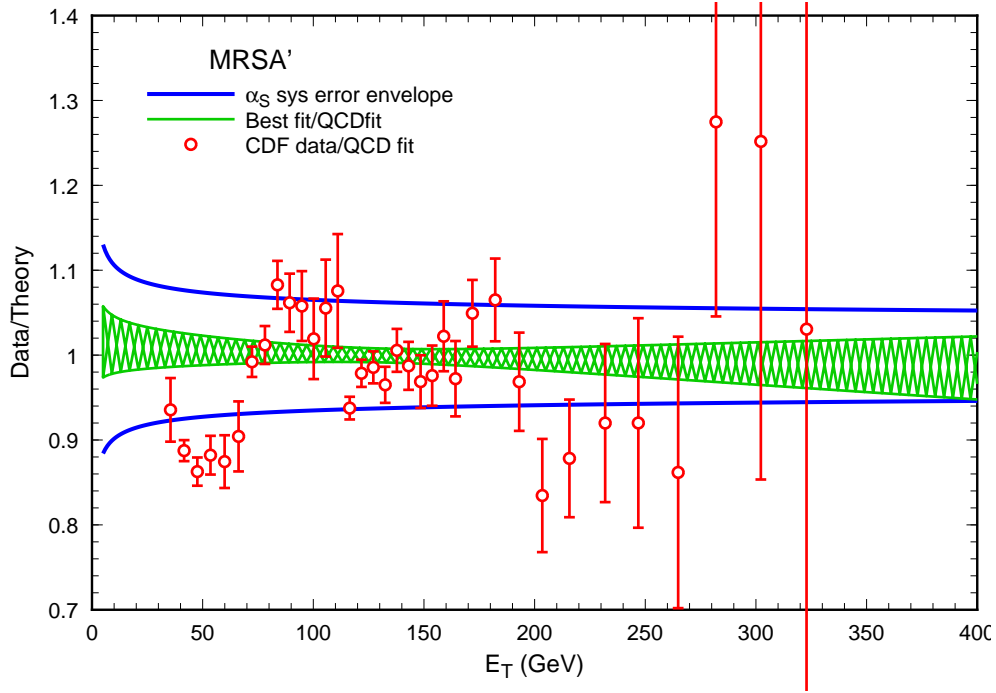}
\caption[]{Ratio of the published CDF data and next-to-leading order
prediction evaluated using $\al^{(1)}(E_T)$ from the QCD- and

Best-fits evolved from $M_Z$ to $E_T$ for

the MRSA$^\prime$ parameterisation.

The systematic uncertainty is also shown

in the QCD-fit based procedure.}
\end{figure}

Finally, we can use the measured evolution of $\al$ to calculate the
one-jet inclusive
cross section. The differential distribution is shown in fig.~5,
while the more useful data divided by theory result is shown
in fig.~6.
Both the QCD-fit (including the systematic
error) and the Best-fit for the MRSA$^\prime$ parameterisation
describe the data well.
The prescaling
thresholds and systematic errors are clearly visible.




Note that if we use the measured running $\al$ for other predictions
and
compare to the CDF '88-'89 data set results the common luminosity
error
would cancel because it is parametrized in the measured $\al$.

\section{Conclusions}

In this paper we have made a first study of
the ability of a hadron collider experiment to extract $\al$
and have utilised the unique feature of hadron colliders to
measure $\al$ over
a wide range of momentum transfers.

As an example we examined the one-jet
inclusive transverse energy distribution and used the CDF '88-'89
data with
an integrated luminosity of 4.2 pb$^{-1}$.

There are two main conclusions.  First, the extracted $\al(M_Z)$ was

consistent with the DIS value of $\al$ used as an input in the $Q^2$

evolution of the parton density functions.
In the future, one can extend this method to include simultaneous
variation
of $\al$ in both the PDF's and the hard scattering cross section.
Second, the measured evolution
of $\al$, as function of the momentum transfer in the scattering, was
shown to be consistent with QCD predictions from 30 GeV up to 500
GeV.

The published data  suffers from large systematic

errors. However, the current run at Fermilab should deliver
in excess of 100 pb$^{-1}$
to both the CDF and D\O\ experiments. This should significantly
reduce the
error on the extracted $\al$ and on its running behavior.

Furthermore, the high luminosity
offers other possibilities to measure $\al$ with high precision, for
example
in high momentum $Z$-boson production which requires only the

measurement of the charged lepton momenta.

With the forthcoming main injector program at Fermilab and an
integrated

luminosity well over 1000 pb$^{-1}$, the $\al$ measurements will keep
improving
significantly in the coming years.

Finally, with such a high luminosity it will be possible to measure
the PDF's

at high $Q^2$ and moderate $x$ values with no input from other
experiments.

This, combined with the $\al$-measurement,
will form a precise test of QCD.

As one makes a high statistic

probe of distance scales, hitherto only partially explored,
any deviations from QCD at high momentum transfers should become

apparent and possible shortcomings in the theory should be
identified.
In the long term,  the LHC will be an excellent machine to both
measure
$\al$ up to very high momentum transfers (up to around 5 TeV)

as well as the PDF's at higher $Q^2$ and lower $x$.

\section*{Acknowledgements}

We thank the CDF and D\O\ collaboration for their help in completing
this analysis.
WTG thanks Dr. B. Bardeen and Prof. S. Bethke for many useful
discussions.
EWNG thanks the Fermilab theory
group for its kind hospitality where this work was initiated.

JY expresses his appreciation to Prof. H. Weerts
for many useful discussions

and to the National Science Foundation of the US goverment

for support in this investigation

\end{document}